\documentclass{article}
\PassOptionsToPackage{numbers,sort,compress}{natbib}

\usepackage{amsmath,amssymb}

 \usepackage[final]{neurips_2025_ml4ps}

\usepackage[utf8]{inputenc} 
\usepackage[T1]{fontenc}    
\usepackage{hyperref}       
\usepackage{url}            
\usepackage{booktabs}       
\usepackage{amsfonts}       
\usepackage{nicefrac}       
\usepackage{microtype}      
\usepackage{xcolor}         
\usepackage{comment}
\usepackage{graphicx}

\newcommand{\cola}{\texttt{MG-PICOLA}}
\newcommand{\mgquijote}{\texttt{QUIJOTE-MG}}
\newcommand{\mgcola}{\texttt{MG-NECOLA}}

\definecolor{darkgreen}{rgb}{0, 0.6, 0}
\title{
MG-NECOLA: Fast Neural Emulators for Modified Gravity Cosmologies
}

\author{J. Bayron Orjuela-Quintana \\ 
    Universidad  del Valle \\ 
    Santiago de Cali  760032,  Colombia \\
    \texttt{john.orjuela@correounivalle.edu.co} \\
    \And
    Mauricio Reyes \\
    Michigan Technological University \\
    Houghton, MI 49931, USA \\
    \texttt{mrhurtad@mtu.edu} \\
    \AND
    Elena Giusarma \\
    Michigan Technological University \\
    Houghton, MI 49931, USA \\
    \texttt{egiusarm@mtu.edu} \\
    \And
    Francisco Villaescusa-Navarro \\
    Center for Computational Astrophysics \\
    162 5th Avenue, New York, NY 10010, USA \\
    \texttt{fvillaescusa@flatironinstitute.org} \\
    \And
    Neerav Kaushal \\
    Michigan Technological University \\
    Houghton, MI 49931, USA \\
    \texttt{kaushal@mtu.edu} \\
    \And
    C\'esar A. Valenzuela-Toledo \\
    Universidad  del Valle \\ 
    Santiago de Cali  760032,  Colombia \\
    \texttt{cesar.valenzuela@correounivalle.edu.co} \\
}

\begin{document}

\maketitle

\begin{abstract}

Observations of the large-scale structure (LSS) provide a powerful test of gravity on cosmological scales, but high-resolution N-body simulations of modified gravity (MG) are prohibitively expensive. We present \textbf{\mgcola}, a convolutional neural network that enhances fast \cola~simulations to near–N-body fidelity at a fraction of the cost. \mgcola~reproduces \mgquijote~N-body results in the power spectrum and bispectrum with better than 1\% accuracy down to non-linear scales ($k \simeq 1~h~\mathrm{Mpc}^{-1}$), while reducing computational time by several orders of magnitude. Importantly, although trained only on $f(R)$ models with massless neutrinos, the network generalizes robustly to scenarios with massive neutrinos, preserving accuracy to within $5\%$ at non-linear scales. This combination of precision and robustness establishes \mgcola~as a practical emulator for producing large ensembles of high-fidelity simulations, enabling efficient exploration of modified gravity and beyond-$\Lambda$CDM cosmologies in upcoming surveys.

\end{abstract}

\section{Introduction}

Modern cosmology and astrophysics increasingly rely on accurate theoretical predictions to extract information from the wealth of data provided by current and upcoming galaxy surveys. Statistical measures such as the matter power spectrum, correlation functions, and higher-order statistics encode the imprint of fundamental physics on the large-scale structure (LSS) of the Universe. The prevailing framework, known as the $\Lambda$ Cold Dark Matter ($\Lambda$CDM) model, combines General Relativity (GR) with a Universe dominated by cold dark matter and dark energy. While $\Lambda$CDM has been highly successful in explaining a broad range of observations, growing evidence suggests possible inconsistencies between different cosmological probes~\cite{DESI:2025, Abdalla:2022yfr, DiValentino:2020vvd}. These observational hints, together with the fact that GR has never been directly tested on the largest cosmic scales, provide strong motivation to explore \textit{modified gravity} (MG) as a compelling extension of the standard model. MG theories alter the growth of cosmic structures by introducing additional degrees of freedom that can enhance or suppress gravitational interactions on cosmological scales. The next generation of surveys will deliver unprecedentedly precise measurements of LSS, creating a unique opportunity to test these alternatives to GR and potentially uncover new physics.

To exploit this opportunity, fast and accurate theoretical predictions are essential. High-resolution $N$-body simulations provide the most reliable predictions for non-linear structure formation, but their computational cost makes them impractical for exploring the vast parameter spaces of MG cosmologies. Approximate methods, such as the \textit{COmoving Lagrangian Acceleration} (COLA)~\cite{cola} approach, are far more efficient and reproduce large-scale clustering well, but their accuracy deteriorates on non-linear scales where MG effects are most pronounced. We address this challenge with \mgcola, a deep learning framework based on convolutional neural networks (CNNs) that maps fast COLA simulations into high-fidelity $N$-body–like realizations. Trained on the \mgquijote~suite~\cite{mgquijote, Villaescusa_Navarro_2020} of extended cosmologies, \mgcola~recovers key summary statistics such as the matter power spectrum down to $k \simeq 1~h~\mathrm{Mpc}^{-1}$ while remaining orders of magnitude faster than full $N$-body runs. This enables efficient exploration of MG scenarios and provides a scalable tool for large-scale structure analyses in upcoming surveys.

\section{Method}

In the standard cosmological paradigm, the growth of structure is governed by GR, where matter density perturbations source the Newtonian potential. MG theories extend this framework by introducing additional degrees of freedom that rescale the gravitational strength in a scale- and time-dependent way~\cite{Clifton:2011jh, CANTATA:2021asi}. A widely studied example is the Hu–Sawicki $f(R)$ model~\cite{Hu:2007pj}, where a scalar degree of freedom mediates an additional force, leaving characteristic signatures in the matter power spectrum at non-linear scales~\cite{DeFelice:2010aj, Orjuela-Quintana:2023zjm}. The strength of this force is controlled by the parameter $f_{R_0}$, with observationally allowed values typically satisfying $|f_{R_0}| < 10^{-5}$~\cite{Basilakos:2013nfa, Cataneo:2014kaa}. In this work, we focus on $f(R)$ gravity as a benchmark to evaluate our deep learning framework, \mgcola.

\subsection {Simulation Data:}

Our framework requires both high-fidelity references and fast approximate simulations. As ground truth, we use the \mgquijote~suite~\cite{mgquijote}, a collection of large cosmological N-body simulations including extensions beyond $\Lambda$CDM. These runs evolve $512^3$ CDM particles in a periodic box of $1000~{\rm Mpc}~h^{-1}$ from redshift $z=127$ to $z=0$. In this context, each particle’s trajectory is described by its displacement, defined as $\Delta = \mathbf{x}_f - \mathbf{x}_i$, where $\mathbf{x}_i$ are the initial grid positions and $\mathbf{x}_f$ the final positions at the present epoch. With thousands of timesteps, N-body simulations provide accurate non-linear predictions but are computationally prohibitive for systematic parameter exploration.

For fast approximations, we generate simulations with the publicly available \cola~code~\cite{mgpicola}, which achieves efficiency by evolving large-scale modes with second-order Lagrangian perturbation theory (2LPT) while treating smaller scales with N-body dynamics~\cite{Bernardeau_2002}. This reduces runtime by more than an order of magnitude, but accuracy deteriorates in the non-linear regime - precisely where MG effects are most relevant.

\subsection {Training and Evaluation:}

\mgcola~is a CNN designed to refine \cola~simulations toward N-body accuracy. The network takes as input the \cola~displacement field, $\Delta_{\rm MG\text{-}COLA}$, and is trained to predict the residual relative to the high-fidelity \mgquijote~reference:
$
  \Delta_{\rm res} = \Delta_{\rm QUIJOTE\text{-}MG} - \Delta_{\rm MG\text{-}COLA}~.
$
The corrected displacement field is then obtained as $\Delta_{\rm MG\text{-}COLA} + \Delta_{\rm res}$, providing an enhanced approximation that closely reproduces the N-body result at a fraction of the computational cost.

Each \cola~run is initialized with the same random seeds and cosmological parameters as its \mgquijote~counterpart, ensuring particle-by-particle correspondence between the two simulations. The dataset consists of 100 paired simulations with the modified gravity parameter fixed at $f_{R_0} = 5 \times 10^{-7}$, while the cosmological parameters are held constant at \mbox{$(\Omega_m, \Omega_b, h, n_s, \sigma_8) = (0.3175, 0.049, 0.6711, 0.9624, 0.8340)$.}

Performance is evaluated against three references: (1) \cola, our fast-simulation benchmark; (2) \mgquijote, the high-fidelity N-body target; and (3) \mgcola, our CNN-enhanced predictions. Assessment is carried out at both the particle level and through cosmological summary statistics, with particular emphasis on the matter power spectrum and bispectrum at non-linear scales.

\subsection{Network Architecture:}
\label{sec:architecture}

Our network design follows the V-Net--based convolutional architecture introduced in the \texttt{NECOLA} framework \cite{NECOLA, Kaushal2}, originally developed for $\Lambda$CDM simulations. While the core structure is similar, this work makes two key advances: (i) it extends the framework to modified gravity cosmologies, where non-linear effects are stronger and more scale-dependent, and (ii) it introduces an adapted loss function that balances accuracy in both Lagrangian and Eulerian displacement fields, along with a gradient-matching term to capture small-scale features. Together, these modifications enable \mgcola~to recover the characteristic signatures of MG models with high fidelity, going beyond the original $\Lambda$CDM application. The novelty thus lies not in the architectural form but in the integration of MG-specific priors into the learning objective and its application to non-standard gravity simulations.

The V-Net-based architecture~\citep{milletari2016vnet, He:2018ggn, Berger:2018aey, deoliveira2020fast} follows a symmetric encoder–decoder (``U-shaped'') design with residual connections. The model consists of two downsampling and two upsampling stages, each comprising residual blocks of two $3^3$ convolutions with $1^3$ shortcut paths. Batch normalization is applied after every convolution except for the first and last two, and leaky ReLU activations (slope $0.01$) follow each normalization as well as the first and penultimate convolutions. The hidden layers use 64 channels, the input and output layers have three channels (corresponding to displacement components), and the concatenated layers use 128 channels.

Residual learning is applied by adding the input displacement field, $\Delta_{\rm MG\text{-}COLA}$, directly to the network output, so that the model learns only the correction toward the target $\Delta_{\rm QUIJOTE\text{-}MG}$. Downsampling and upsampling are implemented using $3^3$ convolutions and $2^3$ transposed convolutions with stride 2. All $3^3$ convolutions are unpadded to preserve translational equivariance; encoder features are therefore cropped before concatenation with decoder layers. The training loss combines three contributions:
\begin{equation}
\label{Eq: Loss Function}
    \mathcal{L} \equiv w_\mathrm{lag}\,\mathcal{L}_\mathrm{lag} 
    + w_\mathrm{eul}\,\mathcal{L}_\mathrm{eul} 
    + w_\mathrm{grad}\,\mathcal{L}_\mathrm{grad},
\end{equation}
where $\mathcal{L}_\mathrm{lag}$ measures the root-mean-squared error (RMSE) on the displacement field, $\mathcal{L}_\mathrm{eul}$ computes the RMSE on the overdensity field, and $\mathcal{L}_\mathrm{grad}$ evaluates the RMSE on the displacement field gradient. This gradient term penalizes both overly smooth and excessively abrupt displacement predictions, ensuring better recovery of small-scale structure particularly important in MG scenarios. We set $(w_{\mathrm{lag}}, w_{\mathrm{eul}}, w_{\mathrm{grad}})=(1.0, 3.0, 2.0)$ so that the three terms contribute comparably in the validation set. This loss formulation represents a significant departure from the \texttt{NECOLA} approach, which used $\mathcal{L}_\mathrm{NECOLA} \equiv \log(\mathcal{L}_\mathrm{eul}\mathcal{L}_\mathrm{lag}^\lambda)$ with hyperparameter $\lambda$, focusing primarily on particle number density and displacements in $\Lambda$CDM. Our three-term loss reflects the need to capture MG-specific small-scale signatures with higher accuracy, as demonstrated in our ablation studies (see bottom left panel of Fig.~\ref{fig:results}).

Given the large data size ($3 \times 512^3$), inputs are split into subcubes of $3 \times 128^3$ (250~Mpc$~h^{-1}$ per side). To preserve translational equivariance, no padding is applied in the $3^3$ convolutions; instead, inputs are periodically padded with 20 voxels per side, yielding effective inputs of $3 \times 164^3$. Data augmentation includes random rotations and parity transformations, ensuring equivariance of the displacement field.  

Training uses Adam optimizer,~\cite{adam}, with a learning rate $10^{-4}$, $\beta_1=0.9$, $\beta_2=0.999$, and a learning rate halving after three stagnant validation epochs. The model is trained for 100 epochs using 70\% of the simulations, with 20\% reserved for validation and 10\% for testing.

\section{Results}
\label{Sec: Results}

\begin{figure}
\centering
 \includegraphics[width=0.5\textwidth]{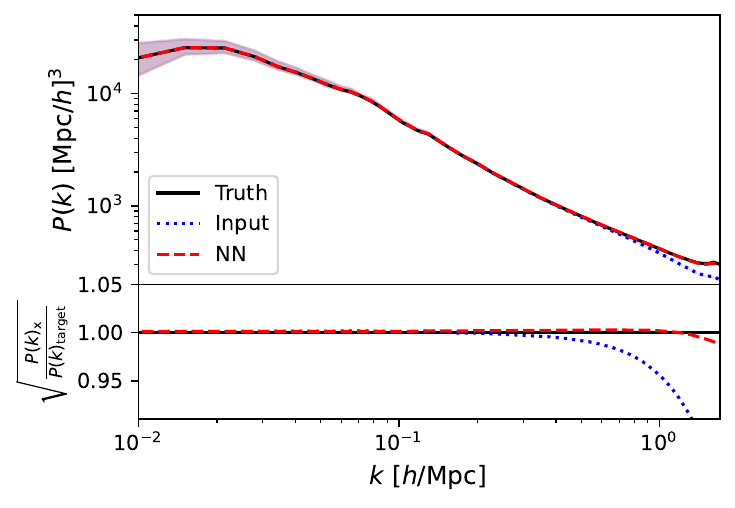} \hfill
 \includegraphics[width=0.45\textwidth]{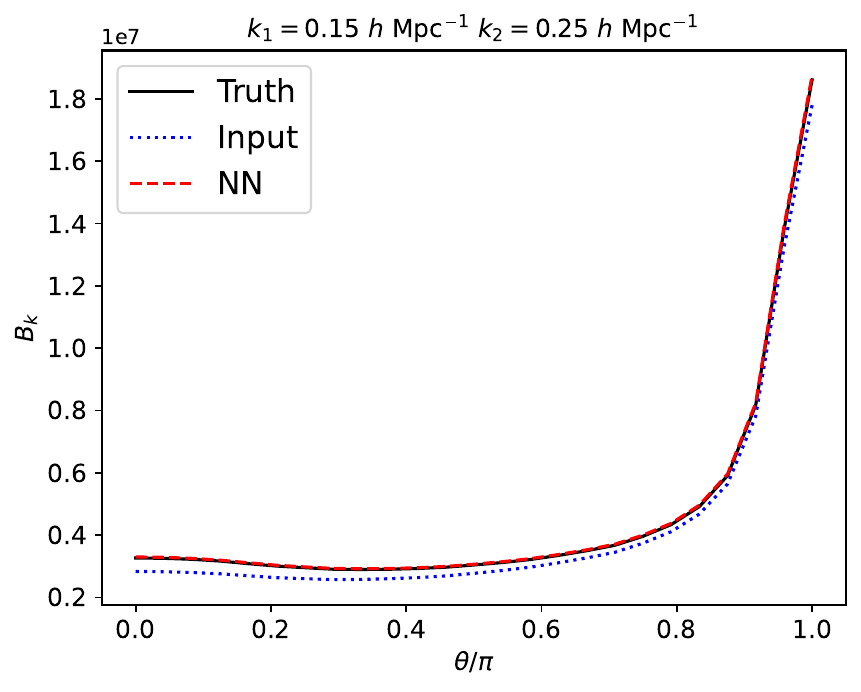}  \\ 
 \includegraphics[width=0.47\textwidth]{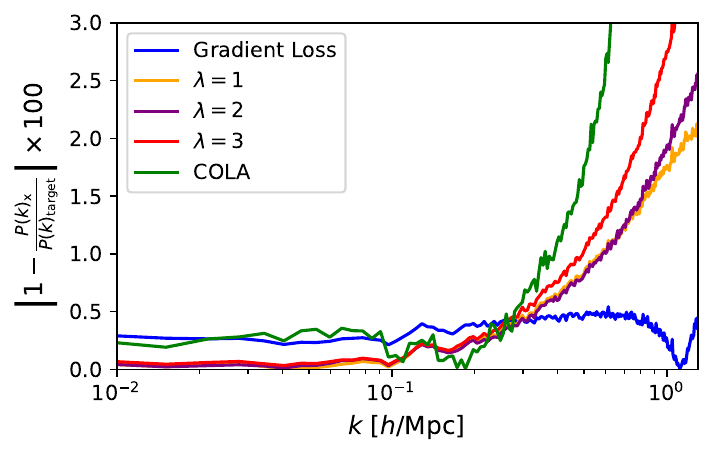} \hfill
 \includegraphics[width=0.5\textwidth]{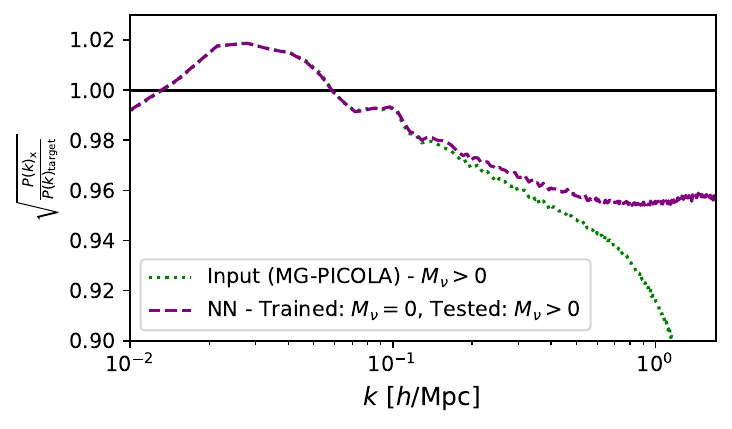}
\caption{Performance of \mgcola~(NN) compared to \cola~(input benchmark) and \mgquijote~(truth). Top left: power spectrum (top) and transfer function (bottom). Top right: bispectrum. Bottom left: ablation study comparing our loss function (Eq.~\eqref{Eq: Loss Function}, blue) with \texttt{NECOLA} loss variants, demonstrating improved accuracy on non-linear scales. Bottom right: transfer function for the generalization to massive neutrinos after training on $M_\nu=0$. \mgcola~(red dashed) substantially improves upon \cola~(blue dotted), recovering results in much closer agreement with the \mgquijote~reference (black solid).}
\label{fig:results}
\end{figure}

Figure~\ref{fig:results} summarizes the performance of \mgcola~on the test set of 10 realizations, evaluated with standard large-scale structure statistics. The top left and right panels show the matter power spectrum, $P(k)$, and the bispectrum, $B(k)$. The power spectrum quantifies the two-point clustering of matter fluctuations as a function of scale, while the bispectrum captures three-point correlations in Fourier space and provides sensitivity to non-linear effects. For the bispectrum, we present the configuration with $k_1=0.15~h~{\rm Mpc}^{-1}$ and $k_2=0.25~h~{\rm Mpc}^{-1}$ as a function of the angle $\theta$ between $k_1$ and $k_2$.

The lower panels show the transfer function, $T(k)$, defined as the square root of the ratio between the predicted and target power spectra, and the fractional percentage accuracy, $\%\mathrm{Acc}$, for one test realization. Values of $T(k)=1$ and $\%\mathrm{Acc} = 0$ indicate perfect recovery of the N-body clustering amplitude.

Across both the power spectrum and bispectrum, \mgcola~(red dashed) provides substantially improved agreement with the \mgquijote~N-body reference (black) compared to \cola~(blue dotted), consistently reducing discrepancies from large to deeply non-linear small scales. The transfer function further demonstrates that \mgcola~effectively corrects the scale-dependent amplitude biases present in \cola, yielding predictions that match the N-body results to better than 1\% accuracy across the full range of probed modes. Furthermore, the bottom left panel of Fig.~\ref{fig:results} shows a small ablation experiment, where it is patented that removing the gradient-matching term ($\mathcal{L}_\mathrm{grad}$) reverting to the original \texttt{NECOLA} loss significantly degrades performance on non-linear scales, particularly in the $k > 0.3~h~\mathrm{Mpc}^{-1}$ regime where MG effects become pronounced. This confirms that our proposed loss formulation, with its explicit emphasis on small-scale structure through gradient matching, provides tangible benefits for capturing MG signatures. These results confirm that \mgcola~can successfully enhance fast \cola~simulations, recovering the accuracy of \mgquijote~N-body runs, particularly at non-linear scales where modified gravity signatures are most significant.

To further assess robustness, we evaluated \mgcola~on $f(R)$ cosmologies with massive neutrinos, despite training only on massless cases. Figure~\ref{fig:results} (bottom right) shows the transfer function for $f_{R_0} = -5.42 \times 10^{-5}$ with $M_\nu = 0.6152~\mathrm{eV}$. Remarkably, \mgcola~(purple dashed) continues to provide accurate enhancements over \cola~(green dotted), closely following the \mgquijote~N-body reference across both large and small scales. Averaged across all modes, \mgcola~reduces \cola~mean error from 25\% down to 9\%, while maintaining absolute deviations below $5\%$ even at the most non-linear scales ($k >0.3~h~\mathrm{Mpc}^{-1}$). These tests are particularly significant because they involve parameters unseen during training, demonstrating that the network has learned transferable features of non-linear structure formation, capturing the interplay of modified gravity and massive neutrinos without explicit exposure.

A typical \mgquijote~N-body simulation requires approximately 500 CPU-hours, or $10^6$ CPU seconds to complete,  while a single \cola~run takes only $10^3$ CPU seconds. Once trained, \mgcola~adds just a lightweight enhancement stage, processing each realization in approximately 180 seconds on a single GPU. Thus, \mgcola~achieves orders-of-magnitude speed-up over N-body runs while retaining near-\mgquijote~accuracy, enabling large ensembles of high-fidelity MG simulations for surveys pipelines.

\section{Conclusions}

In this work, we introduced \mgcola, a convolutional neural network that upgrades fast \cola~simulations to the accuracy of \mgquijote~N-body runs with significantly reduced computational cost. Using standard large-scale structure statistics, we showed that \mgcola~reproduces N-body results with better than 1\% accuracy down to non-linear scales of $k \sim 1~h~{\rm Mpc}^{-1}$, significantly outperforming \cola. Moreover, although trained only on a single $f_{R_0}$ value with massless neutrinos, the network generalizes robustly to cosmologies with massive neutrinos and varying cosmological parameters, demonstrating strong transferability to unseen physics.

Looking ahead, the next step is to extend \mgcola~to a broader family of non-standard cosmological models. In particular, we aim to incorporate diverse values of $f_{R_0}$ together with varying neutrino masses, ultimately building a framework capable of mapping across the space of modified gravity and beyond-$\Lambda$CDM scenarios. Such a tool would allow systematic exploration of cosmological signatures across a wide range of theories, providing an essential resource for interpreting data from the next generation of large-scale structure surveys.


\bibliographystyle{plainnat}
\bibliography{references}

\begin{thebibliography}{22}
\providecommand{\natexlab}[1]{#1}
\providecommand{\url}[1]{\texttt{#1}}
\expandafter\ifx\csname urlstyle\endcsname\relax
  \providecommand{\doi}[1]{doi: #1}\else
  \providecommand{\doi}{doi: \begingroup \urlstyle{rm}\Url}\fi

\bibitem[Abdalla et~al.(2022)]{Abdalla:2022yfr}
Elcio Abdalla et~al.
\newblock {Cosmology intertwined: A review of the particle physics,
  astrophysics, and cosmology associated with the cosmological tensions and
  anomalies}.
\newblock \emph{JHEAp}, 34:\penalty0 49--211, 2022.
\newblock \doi{10.1016/j.jheap.2022.04.002}.

\bibitem[Abdul~Karim et~al.(2025)]{DESI:2025}
M.~Abdul~Karim et~al.
\newblock {DESI DR2 Results II: Measurements of Baryon Acoustic Oscillations
  and Cosmological Constraints}.
\newblock 3 2025.

\bibitem[Akrami et~al.(2021)]{CANTATA:2021asi}
Yashar Akrami et~al.
\newblock \emph{{Modified Gravity and Cosmology. An Update by the CANTATA
  Network}}.
\newblock Springer, 2021.
\newblock ISBN 978-3-030-83714-3, 978-3-030-83717-4, 978-3-030-83715-0.
\newblock \doi{10.1007/978-3-030-83715-0}.

\bibitem[{Alves de Oliveira} et~al.(2020){Alves de Oliveira}, {Li},
  {Villaescusa-Navarro}, {Ho}, and {Spergel}]{deoliveira2020fast}
Renan {Alves de Oliveira}, Yin {Li}, Francisco {Villaescusa-Navarro}, Shirley
  {Ho}, and David~N. {Spergel}.
\newblock {Fast and Accurate Non-Linear Predictions of Universes with Deep
  Learning}.
\newblock \emph{arXiv e-prints}, art. arXiv:2012.00240, November 2020.

\bibitem[Baldi and Villaescusa-Navarro(2025)]{mgquijote}
Marco Baldi and Francisco Villaescusa-Navarro.
\newblock Quijote-mg simulations.
\newblock In preparation, 2025.

\bibitem[Basilakos et~al.(2013)Basilakos, Nesseris, and
  Perivolaropoulos]{Basilakos:2013nfa}
Spyros Basilakos, Savvas Nesseris, and Leandros Perivolaropoulos.
\newblock {Observational constraints on viable f(R) parametrizations with
  geometrical and dynamical probes}.
\newblock \emph{Phys. Rev. D}, 87\penalty0 (12):\penalty0 123529, 2013.
\newblock \doi{10.1103/PhysRevD.87.123529}.

\bibitem[Berger and Stein(2019)]{Berger:2018aey}
Philippe Berger and George Stein.
\newblock {A volumetric deep Convolutional Neural Network for simulation of
  mock dark matter halo catalogues}.
\newblock \emph{Mon. Not. Roy. Astron. Soc.}, 482\penalty0 (3):\penalty0
  2861--2871, 2019.
\newblock \doi{10.1093/mnras/sty2949}.

\bibitem[Bernardeau et~al.(2002)Bernardeau, Colombi, Gaztañaga, and
  Scoccimarro]{Bernardeau_2002}
F.~Bernardeau, S.~Colombi, E.~Gaztañaga, and R.~Scoccimarro.
\newblock Large-scale structure of the universe and cosmological perturbation
  theory.
\newblock \emph{Physics Reports}, 367\penalty0 (1-3):\penalty0 1–248, Sep
  2002.
\newblock ISSN 0370-1573.
\newblock \doi{10.1016/s0370-1573(02)00135-7}.
\newblock URL \url{http://dx.doi.org/10.1016/S0370-1573(02)00135-7}.

\bibitem[Cataneo et~al.(2015)Cataneo, Rapetti, Schmidt, Mantz, Allen,
  Applegate, Kelly, von~der Linden, and Morris]{Cataneo:2014kaa}
Matteo Cataneo, David Rapetti, Fabian Schmidt, Adam~B. Mantz, Steven~W. Allen,
  Douglas~E. Applegate, Patrick~L. Kelly, Anja von~der Linden, and R.~Glenn
  Morris.
\newblock {New constraints on $f(R)$ gravity from clusters of galaxies}.
\newblock \emph{Phys. Rev. D}, 92\penalty0 (4):\penalty0 044009, 2015.
\newblock \doi{10.1103/PhysRevD.92.044009}.

\bibitem[Clifton et~al.(2012)Clifton, Ferreira, Padilla, and
  Skordis]{Clifton:2011jh}
Timothy Clifton, Pedro~G. Ferreira, Antonio Padilla, and Constantinos Skordis.
\newblock {Modified Gravity and Cosmology}.
\newblock \emph{Phys. Rept.}, 513:\penalty0 1--189, 2012.
\newblock \doi{10.1016/j.physrep.2012.01.001}.

\bibitem[De~Felice and Tsujikawa(2010)]{DeFelice:2010aj}
Antonio De~Felice and Shinji Tsujikawa.
\newblock {f(R) theories}.
\newblock \emph{Living Rev. Rel.}, 13:\penalty0 3, 2010.
\newblock \doi{10.12942/lrr-2010-3}.

\bibitem[Di~Valentino et~al.(2021)]{DiValentino:2020vvd}
Eleonora Di~Valentino et~al.
\newblock {Cosmology Intertwined III: $f \sigma_8$ and $S_8$}.
\newblock \emph{Astropart. Phys.}, 131:\penalty0 102604, 2021.
\newblock \doi{10.1016/j.astropartphys.2021.102604}.

\bibitem[He et~al.(2019)He, Li, Feng, Ho, Ravanbakhsh, Chen, and
  P{\'o}czos]{He:2018ggn}
Siyu He, Yin Li, Yu~Feng, Shirley Ho, Siamak Ravanbakhsh, Wei Chen, and
  Barnab{\'a}s P{\'o}czos.
\newblock {Learning to Predict the Cosmological Structure Formation}.
\newblock \emph{Proc. Nat. Acad. Sci.}, 116\penalty0 (28):\penalty0
  13825--13832, 2019.
\newblock \doi{10.1073/pnas.1821458116}.

\bibitem[Hu and Sawicki(2007)]{Hu:2007pj}
Wayne Hu and Ignacy Sawicki.
\newblock {A Parameterized Post-Friedmann Framework for Modified Gravity}.
\newblock \emph{Phys. Rev. D}, 76:\penalty0 104043, 2007.
\newblock \doi{10.1103/PhysRevD.76.104043}.

\bibitem[Kaushal et~al.(2021)Kaushal, Villaescusa-Navarro, Giusarma, Li, Hawry,
  and Reyes]{Kaushal2}
Neerav Kaushal, Francisco Villaescusa-Navarro, Elena Giusarma, Yin Li, Conner
  Hawry, and Mauricio Reyes.
\newblock Necola: Toward a universal field-level cosmological emulator.
\newblock In \emph{{35th conference on Neural Information Processing Systems}},
  12 2021.
\newblock URL \url{https://ml4physicalsciences.github.io/2021/}.

\bibitem[Kaushal et~al.(2022)Kaushal, Villaescusa-Navarro, Giusarma, Li, Hawry,
  and Reyes]{NECOLA}
Neerav Kaushal, Francisco Villaescusa-Navarro, Elena Giusarma, Yin Li, Conner
  Hawry, and Mauricio Reyes.
\newblock {NECOLA}: Toward a universal field-level cosmological emulator.
\newblock \emph{The Astrophysical Journal}, 930\penalty0 (2):\penalty0 115, May
  2022.
\newblock \doi{10.3847/1538-4357/ac5c4a}.
\newblock URL \url{https://doi.org/10.3847/1538-4357/ac5c4a}.

\bibitem[Kingma and Ba(2017)]{adam}
Diederik~P. Kingma and Jimmy Ba.
\newblock Adam: A method for stochastic optimization, 2017.

\bibitem[Milletari et~al.(2016)Milletari, Navab, and Ahmadi]{milletari2016vnet}
Fausto Milletari, Nassir Navab, and Seyed-Ahmad Ahmadi.
\newblock V-net: Fully convolutional neural networks for volumetric medical
  image segmentation, 2016.

\bibitem[Orjuela-Quintana and Nesseris(2023)]{Orjuela-Quintana:2023zjm}
J.~Bayron Orjuela-Quintana and Savvas Nesseris.
\newblock {Tracking the validity of the quasi-static and sub-horizon
  approximations in modified gravity}.
\newblock \emph{JCAP}, 08:\penalty0 019, 2023.
\newblock \doi{10.1088/1475-7516/2023/08/019}.

\bibitem[Tassev et~al.(2013)Tassev, Zaldarriaga, and Eisenstein]{cola}
Svetlin Tassev, Matias Zaldarriaga, and Daniel~J Eisenstein.
\newblock Solving large scale structure in ten easy steps with {COLA}.
\newblock \emph{Journal of Cosmology and Astroparticle Physics}, 2013\penalty0
  (06):\penalty0 036--036, jun 2013.
\newblock \doi{10.1088/1475-7516/2013/06/036}.
\newblock URL \url{https://doi.org/10.1088}.

\bibitem[Villaescusa-Navarro et~al.(2020)Villaescusa-Navarro, Hahn, Massara,
  Banerjee, Delgado, et~al.]{Villaescusa_Navarro_2020}
Francisco Villaescusa-Navarro, ChangHoon Hahn, Elena Massara, Arka Banerjee,
  Ana~Maria Delgado, et~al.
\newblock The quijote simulations.
\newblock \emph{The Astrophysical Journal Supplement Series}, 250\penalty0
  (1):\penalty0 2, August 2020.
\newblock \doi{10.3847/1538-4365/ab9d82}.
\newblock URL \url{https://doi.org/10.3847/1538-4365/ab9d82}.

\bibitem[{Wright} et~al.(2017){Wright}, {Winther}, and {Koyama}]{mgpicola}
Bill~S. {Wright}, Hans~A. {Winther}, and Kazuya {Koyama}.
\newblock {COLA with massive neutrinos}.
\newblock 2017\penalty0 (10):\penalty0 054, October 2017.
\newblock \doi{10.1088/1475-7516/2017/10/054}.

\end{thebibliography}

\end{document}